\documentclass[final,twoside,11pt]{entics} 
\usepackage{enticsmacro}
\usepackage{macros}
\usepackage[all]{xy}\NoCompileMatrices
 \pdfoutput=1 

\def\conf{MFPS 2023} 	
\def\myconf{mfps39}
\volume{3}			
\def\volu{3}			
\def\papno{8}				
\def\lastname{Fiore, Galal, Jafarrahmani} 
\def\runtitle{Fixpoints constructions in focused orthogonality models of linear logic} 
\def\mydoi{12302}
\begin{document}
\begin{frontmatter}
  \title{Fixpoint Constructions in Focused Orthogonality Models of Linear Logic} 						
  \author{Marcelo Fiore\thanksref{a}\thanksref{c}\thanksref{email1}}	
   \author{Zeinab Galal\thanksref{b}\thanksref{c}\thanksref{d}\thanksref{email2}}		  
    \author{Farzad Jafarrahmani\thanksref{b}\thanksref{d}\thanksref{email3}}		
   \address[a]{University of Cambridge}  							
   \thanks[c]{Research partially supported by EPSRC grant EP/V002309/1.}
    \thanks[d]{Research partially supported by the RealiSe Emergence project.}
   \thanks[email1]{Email: \href{mailto:marcelo.fiore@cl.cam.ac.uk} {\texttt{\normalshape
        marcelo.fiore@cl.cam.ac.uk}}} 
  \address[b]{Sorbonne University} 
  \thanks[email2]{Email:  \href{mailto:zeinab.galal@lip6.fr} {\texttt{\normalshape
       zeinab.galal@lip6.fr}}}
    \thanks[email3]{Email:  \href{mailto:farzad.jafarrahmani@lip6.fr} {\texttt{\normalshape
   			farzad.jafarrahmani@lip6.fr}}}
   		
\begin{abstract} 
Orthogonality is a notion based on the duality between programs and their
environments used to determine when they can be safely combined.  For
instance, it is a powerful tool to establish termination properties in
classical formal systems.  
It was given a general treatment with the concept of orthogonality category,
of which numerous models of linear logic are instances, 
by
Hyland and Schalk. 
This paper considers the subclass of focused orthogonalities.

We develop a theory of fixpoint constructions in focused orthogonality
categories.
Central results are lifting theorems for initial algebras and final
coalgebras.
These crucially hinge on the insight that focused orthogonality categories are
relational fibrations.  
The theory provides an axiomatic categorical framework for models of linear
logic with least and greatest fixpoints of types.

We further investigate domain-theoretic settings, showing how to lift bifree
algebras, used to solve mixed-variance recursive type equations, to focused
orthogonality categories. 
\end{abstract}

\end{frontmatter}

\section*{Introduction}\label{sec:intro}
\subsection*{Linear logic with fixpoints}

Propositional linear logic lacks datatypes with iteration or recursion
principles.  This is usually remedied by extending it to the second order,
thus defining a logical system in which Girard's System $\SF$~\cite{Girard89}
can be embedded.  Even if such a system is very expressive in terms of
computable functions, its algorithmic expressiveness is poor: for instance, it
is not possible to write a term

that computes the predecessor function in one (or a uniformly bounded) number
of reduction steps. 

Girard first considered an extension of linear logic with fixpoints of
formulas in an unpublished note~\cite{Girard92}.  However, the first
comprehensive proof-theoretic investigation of such a system was given by
Baelde~\cite{Baelde12} who introduced and studied $\MUMALL$, an extension of
multiplicative additive linear logic with induction and coinduction
principles, with motivations coming from proof-search and system verification.
Linear logic exponentials were not considered in $\MUMALL$; they could be
somewhat encoded with inductive and coinductive types though without their
denotational interpretation satisfying the required Seely isomorphisms.

Ehrhard and Jafarrahmani~\cite{EJ2021} introduced a system $\MULL$ extending
$\MUMALL$ with exponentials and studied it from the Curry-Howard-Lambek
perspective.  Their notion of categorical model of $\MULL$ is an extension of
the standard notion of Seely category for classical linear logic with suitable
initial algebras and final coalgebras.  Specifically, they presented a
totality model of $\MULL$ that is an instance of a general categorical
construction developed in this paper.  In the totality model, least and
greatest fixpoints are calculated by lifting initial algebras and final
coalgebras from the relational model.  Here, by viewing it as a special case
of a \emph{focused orthogonality} construction, we are able to develop a
general methodology for constructing models of linear logic with fixpoints.

\subsection*{Orthogonality and glueing for models of linear logic}

Logical relations~\cite{tait1967intensional,plotkin1973lambda,Statman} are by
now a standard tool in the theory of programming languages to certify program
properties that cannot be obtained by naive induction arguments.  The basic
idea is to associate to each type a predicate that is preserved by the
operations on the type.  Such predicates depend on the program property that
one is interested in proving (termination, type safety, parametricity, etc.)
and their use provides a powerful proof method.

Orthogonality methods originate from the semantics of linear logic and are
particularly well-suited for languages modelling classical
negation~\cite{GIRARDLL}. 
The general principle is to restrict attention to pairs of terms and contexts
in a \emph{pole} $\Perp \: \subseteq  \mathrm{Terms} \times \mathrm{Contexts}$ 
that contains \emph{correct computations}.  For a set of terms 
$T \subseteq \mathrm{Terms}$, one can then consider the set of all contexts
$T^\orth \subseteq \mathrm{Contexts}$ that yield a correct computation when
combined with any term in $T$.  
Dually, for a set of contexts $C \subseteq \mathrm{Contexts}$, one can
consider the set of all terms $C^\orth\subseteq\mathrm{Terms}$ that yield a
correct computation when combined with any context in $C$.
These constructions yield a duality between subsets of terms and subsets of
contexts, and one associates to each type a subset of terms $T$ that is equal
to its double dual $T^\dorth$.
Such dualities between terms and environments (or player and opponent) form the basis of game semantics~\cite{Hyland1997} and of Krivine's classical
realizability~\cite{krivine2009realizability}.

Logical relations have a categorical abstraction given by Artin-Wraith 
glueing~\cite{Wraith}, while orthogonality constructions are obtained via
Hyland-Schalk double glueing~\cite{GlueingHylandSchalk}.  
Here, we will be particularly interested in a well-behaved subclass of
orthogonality categories arising from poles and referred to as 
\emph{focused orthogonality categories}.  
These, we will recast as relational fibrations and therefrom develop a general
categorical theory that lifts initial algebras and final coalgebras to focused
orthogonality categories, and therefore provides models of linear logic with
least and greatest fixpoints.

\subsection*{Structure of the paper}
\begin{itemize}
\item 
  We start by recalling the notion of orthogonality category by Hyland and
  Schalk in Section~\ref{sec:prelim}.

\item 
  In Section~\ref{Sec:FixConsRelFib}, we develop a theory of fixpoint
  constructions for relational fibrations by lifting initial algebras and
  final coalgebras to the Grothendieck category of a relational fibration.

\item 
  We show in Section~\ref{sec:focusedFib} that focused orthogonality models
  are instances of relational fibrations.  This provides us with a general
  categorical construction to obtain models of linear logic with least and
  greatest fixpoints.  A variety of examples is considered in
  Section~\ref{Sec:Models}.

\item 
  Finally, in Section~\ref{sec:enrichedterms}, 
  we show how to lift bifree algebras to focused orthogonality categories in
  an axiomatic domain-theoretic setting. 
\end{itemize}

\section{Preliminaries on orthogonality categories}\label{sec:prelim}

From a categorical viewpoint, logical relations can be presented using glueing
constructions, also called Artin-Wraith glueing, sconing, or Freyd
covering~\cite{artin1973theorie,Wraith,freyd1990categories}.  These allow the
lifting of monoidal (or cartesian) closed structure to glueing categories. Orthogonality methods fit into the more general framework of double-glueing
constructions by Hyland, Schalk, and
Tan~\cite{tan1998full,GlueingHylandSchalk} which is tailored to
$\star$-autonomous categories.  The general idea is to associate two
predicates with each type: one for the type and another one for its dual. 

For a $\star$-autonomous category $\cat{C}$ with monoidal units $\One$ and
$\bot$, an \emph{orthogonality relation} $\perp$ is a family of subsets 
\[
\perp_c \enspace \subseteq \enspace \cat{C}(\One, c) \times \cat{C}(c, \bot)
\]
indexed by objects $c \in \cat{C}$ and verifying some compatibility conditions
with respect to the linear logic structure~\cite{GlueingHylandSchalk}. 
For a subset $X \subseteq \cat{C}(\One, c)$, its \emph{orthogonal} 
$X^\orth \subseteq \cat{C}(c, \bot)$ is given by
$X^\orth 
 \eqdef \setof{\, y : c \to \bot \mid \forall\, x \in X.\, x \perp_c y \,}$ 
with the idea that $X^\perp$ contains the environments (or counter-terms) that
yield a correct computation (with respect to the chosen orthogonality
relation) when combined with any term in $X$.  
Dually, for a subset $Y \subseteq \cat{C}(c, \bot)$, its orthogonal 
$Y^\orth \subseteq \cat{C}(\One, c)$ is given by 
$Y^\orth 
 \eqdef \setof{\, x : \One \to c \mid \forall\, y \in Y.\, x \perp_c y \,}$.
We restrict attention subsets of global elements that are double orthogonally
closed and define the \emph{orthogonality category induced by $\{\perp_c\}_{c\in\lscat C}$} to
have objects given by pairs $(c, X)$ with 
$X= X^\dorth \subseteq \cat{C}(\One, c)$ and morphisms $(c, X)\to (d, Y)$
given by morphisms $f :c \to d$ in $\lscat C$ such that: 
\begin{equation}\label{OrthCatCondition}
	\forall\, x \in X.\, f x\in Y 
  \quad \text{ and } \quad 
  \forall\, y \in Y^\perp.\, yf \in X^\orth
  \enspace.
\end{equation}

Provided that some conditions on $\cat{C}$ and the orthogonality relation
$\{\perp_c\}_{c\in\lscat C}$ hold, if $\cat{C}$ is a model of classical linear
logic then so is the induced orthogonality category, and the forgetful functor
preserves the linear logic structure strictly~\cite{GlueingHylandSchalk}. 
In this paper, we will restrict to the special case where the orthogonality
relation arises from a distinguished subset 
$\Perp\: \subseteq \cat{C}(\One, \bot)$, referred to as a pole, as follows: 
\[
  \Perp_c
  \enspace \eqdef \enspace
  \setof{\, 
  (x, y) \in \cat{C}(\One, c) \times \cat{C}(c, \bot) 
  \suchthat 
  y \icomp x \in\: \Perp
  \,}
\enspace.
\] 
Such orthogonality relations are called \emph{focused} and for them the two
conditions in~(\ref{OrthCatCondition}) above are
equivalent~\cite{GlueingHylandSchalk}.  
This property will crucially allow us to subsume focused orthogonality
categories within a fibrational setting and, from the theory of fixpoint
constructions for relational fibrations of the following section, we will
obtain models of linear logic with least and greatest fixpoints.

\section{Fixpoint constructions in relational fibrations}
\label{Sec:FixConsRelFib}
This section develops a general method to lift initial algebras and final
coalgebras form the base category of a relational fibration to its total
category.

\subsection{Relational fibrations}

We start by recalling some basic properties of relational fibrations.

\begin{definition}
A \emph{$\lscat{C}$-indexed poset} is a contravariant functor from a category
$\lscat C$ to the category $\Poset$ of posets and monotone functions between
them.
\end{definition}

For an indexed poset $\R : \cat{C}^\op \to \Poset$, a morphism $f: c\to d$ in
$\lscat C$, and $S\in\R(d)$, it is customary to write $f\upstar(S)$ for
$\R(f)(S)\in\R(c)$.  For $R\in\R(c)$, we moreover write $f: R \supset S$ for
$R\leq f\upstar(S)$.

\begin{definition}
The \emph{Grothendieck category} $\gcat{\R}{\cat{C}}$ of a $\cat{C}$-indexed
poset $\R : \cat{C}^\op \to \Poset$ has objects given by pairs $\Gobj c R$
with $c \in \cat{C}$ and $R \in \R(c)$, and morphisms 
$f : {\Gobj c R} \to \Gobj d S$  given by morphisms $f: c \to d$ in $\cat{C}$
such that $f: R \supset S$ in $\R(c)$. Identities and composition are given as
in $\cat{C}$.
\end{definition}

The forgetful functor $U: \gcat{\R}{\cat{C}} \to \cat{C}$	is a Grothendieck
fibration with partially-ordered fibers.  
Note that, for every $c\in\lscat C$, $R\leq R'$ in $\R(c)$ if and only if
$\id[c]:\Gobj c R\to \Gobj c{R'}$ in $\gcat{\R}{\cat{C}}$.

We refer to $U$ as the \emph{relational fibration} of the $\cat{C}$-indexed
poset $\R$.
The cartesian lifting of $f: c \to d$ in $\lscat C$ with respect to 
$\Gobj d S\in \gcat{\R}{\cat{C}}$ is the morphism 
$f: {\Gobj c {f\upstar S}} \to \Gobj d S$ in $\gcat{\R}{\cat{C}}$.

\begin{definition}
For a $\cat{C}$-indexed poset $\R$, we say that an endofunctor $\lift F$ on
$\gcat{\R}{\cat{C}}$ is a \emph{lifting} of an endofunctor $F$ on $\cat{C}$
whenever the following diagram commutes
	\begin{center}
		\begin{tikzpicture}
			\node (A) at (2,0) {$\cat{C}$};
			\node (B) at (2,1.5) {$\gcat{\R}{\cat{C}}$};
			\node (C) at (0,0) {$\cat{C}$};
			\node (D) at (0,1.5) {$\gcat{\R}{\cat{C}}$};
			
			\draw [->] (C) -- node [below] {$F$} (A);
			\draw [->] (D) -- node [above] {$\lift F$} (B);
			\draw [->] (B) -- node [right] {$U$} (A);
			\draw [->] (D) -- node [left] {$U$} (C);
		\end{tikzpicture}
	\end{center}
\end{definition}
We let $\lift{F}$ be a lifting of $F$ as above for the rest of the section.

For ${\Gobj c R}\in \gcat{\R}{\cat{C}}$, we write $\lift F_c(R)$ for the
element in $\R(Fc)$ given by ${\lift F}\,{\Gobj c R}$; in other words, we let
${\lift F}\,{\Gobj c R} = \Gobj{Fc\,}{\,\lift F_c(R)}$.  
Therefore, for all $f: \Gobj c R \to \Gobj d S$ in $\gcat{\R}{\cat{C}}$, since
$\lift F(f) = F(f)$, we have that 
$\lift F_c(R) \leq (Ff)\upstar\big(\lift F_d(S)\big)$ in $\R(Fc)$.  This has
the following direct consequences.
\begin{lemma}\label{DirectConsequences}
\begin{enumerate}
\item \label{LiftFMonotonicity}
For all $c\in\lscat C$, the function $\lift F_c: \R(c) \to \R(Fc)$ is
monotone.  

\item \label{LiftFOfCartesianLifting}
For all $f: c\to d$ in $\cat{C}$ and $S\in \R(d)$,  
$\lift F_c\big(f\upstar(S)\big) \leq (Ff)\upstar\big(\lift F_d(S)\big)$ in 
$\R(Fc)$. 
\end{enumerate}
\end{lemma}

\subsection{Initial-algebra lifting theorem}

By Lemma~\ref{DirectConsequences}(\ref{LiftFMonotonicity}), every coalgebra
$\gamma: c\to Fc$ induces the monotone operator 
\[
	\R(c) \xrightarrow{\lift F_c} \R(Fc) \xrightarrow{\gamma \upstar} \R(c)\enspace.
\]
A lifting of the $F$-coalgebra
$\gamma$ to an $\lift F$-coalgebra amounts to providing a post fixpoint of
$\gamma\upstar \comp \lift F_c$; that is, an $R\in\R(c)$ such that 
$R \leq \gamma\upstar\big(\lift F_c(R)\big)$.  On the other hand, a lifting of
an $F$-algebra $\delta: Fd \to d$ amounts to providing an $S\in\R(d)$ such
that $\lift F_d(S)\leq\delta\upstar(S)$.

We now consider homomorphisms from a coalgebra $\gamma: c\to Fc$ to an algebra
$\delta: Fd \to d$ as given by morphisms $f: c\to d$ such that the following diagram commutes:
\[\xymatrix{
    Fc \ar[r]^-{Ff} & Fd \ar[d]^-{\delta}
    \\
    c \ar[u]^-\gamma \ar[r]_-f & d
}\]

\begin{lemma} \label{AlgebraKeyLemma}
For a coalgebra $\gamma: c\to Fc$, the least pre-fixpoint
$\lpfp_\gamma\in\R(c)$ of the monotone operator $\gamma\upstar\comp\lift F_c$,
whenever it exists, provides a lifting of $\gamma$ such that for all liftings
$S\in\R(d)$ of an algebra $\delta: Fd\to d$, every homomorphism $c\to d$ from
$\gamma$ to $\delta$ lifts to a morphism 
$\Gobj c {\lpfp_\gamma} \to \Gobj d S$.  
\end{lemma}
\begin{proof}
We have 
$\gamma: \Gobj c{\lpfp_\gamma} \to \Gobj{Fc\,}{\,\lift F_c(\lpfp_\gamma)}$
because $\lpfp_\gamma$ is a fixpoint of $\gamma\upstar\comp\lift F_c$.

Let $\delta: Fd\to d$ and $S\in\R(d)$ be such that 
$\lift F_d(S)\leq\delta\upstar(S)$ and let $f: c \to d$ be an homomorphism from
$\gamma$ to $\delta$.

By Lemma~\ref{DirectConsequences}(\ref{LiftFOfCartesianLifting}) and the
assumption on $S$, we have
\[
\lift F_c\big(f\upstar(S)\big) 
\leq 
(Ff)\upstar\big(\lift F_d(S)\big)
\leq 
(Ff)\upstar\big(\delta\upstar(S)\big) 
\]
and hence
\[
(\gamma\upstar\comp\lift F_c)\big(f\upstar(S)\big)
 \leq 
 \big(\delta\comp(Ff)\comp\gamma\big)\upstar(S)
 = 
 f\upstar(S)
 \enspace;
\]
that is, $f\upstar(S)$ is a pre-fixpoint of $\gamma\upstar\comp\lift F_c$. 
Therefore, $\lpfp_\gamma\leq f\upstar(S)$ as required.
\end{proof}

\begin{theorem}
For an initial $F$-algebra $\alpha: Fa \to a$, if the monotone operator
$(\alpha^{-1})\upstar\comp\lift F_a$ on $\R(a)$ has a least pre-fixpoint
$\lpfp_{\alpha}$ then 
$\alpha: \Gobj{Fa\,}{\,\lift F_a(\lpfp_\alpha)} \to \Gobj a{\lpfp_\alpha}$
is an initial $\lift F$-algebra. 
\end{theorem}
\begin{proof}
For every $\lift F$-algebra 
$\delta: \Gobj{Fd\,}{\,\lift F_d(S)} \to \Gobj d S$ the unique homomorphism
$u(\delta): a \to d$ from $\alpha: Fa \to a$ to $\delta: Fd \to d$ is a
homomorphism from $\alpha^{-1}: a \to Fa$ to $\delta: Fd \to d$ and, by
Lemma~\ref{AlgebraKeyLemma}, it is also the unique homomorphism from 
$\alpha: \Gobj{Fa\,}{\,\lift F_a(\lpfp_\alpha)} \to \Gobj a{\lpfp_\alpha}$ to 
$\delta: \Gobj{Fd\,}{\,\lift F_d(S)} \to \Gobj d S$.
\end{proof}

Let $\alpha: Fa \to a$ be an initial algebra and, for an algebra 
$\delta:Fd\to d$, let $u(\delta):a\to d$ be the unique homomorphism from
$\alpha$ to $\delta$.  In the situation of the theorem, initial algebras
$\alpha$ satisfy the following property: 
\begin{quote} \centering 
  for every algebra $\delta: Fd\to d$ and $S\in\R(d)$, if
  $\delta:\lift F_d(S)\supset S$ then 
  $u(\delta): \lpfp_\alpha \supset S$\enspace.
\end{quote}
This provides an abstract general form of \emph{induction principle}.  Indeed,
in particular, one has:
\begin{quote} \centering 
  for every $S\in\R(a)$, if $\lift F_a(S)\leq \alpha\upstar(S)$ then
  $\lpfp_\alpha \leq S$\enspace.
\end{quote}
As advocated by Hermida and Jacobs~\cite[Definition~3.1]{HJ}, the standard
induction principle is recovered when $\lpfp_\alpha$ is the greatest element
$\top_a$ of $\R(a)$, in which case one has:
\begin{quote} \centering 
  for every algebra $\delta: Fd\to d$ and $S\in\R(d)$, if
  $\delta:\lift F_d(S)\supset S$ then 
  $u(\delta)\upstar(S)=\top_a$
\end{quote}
and, in particular, that:
\begin{quote} \centering 
  for every $S\in\R(a)$, if $\lift F_a(S)\leq \alpha\upstar(S)$ then 
  $S = \top_a$\enspace.
\end{quote}

An \emph{ipo} (inductive poset) is a poset with a least element and joins of
directed subsets.  Such posets are particularly relevant to us here because of
Pataraia's constructive theorem that every monotone endofunction on an ipo has
a least pre-fixpoint~\cite{Pataraia}.

\begin{definition}
A \emph{$\lscat{C}$-indexed ipo} is a $\lscat C$-indexed poset $\R$ such that
$\R(c)$ is an ipo for all $c\in\lscat C$.
\end{definition}

\begin{corollary}
For every $\lscat C$-indexed ipo $\R$, every endofunctor $F$ on $\lscat C$,
and every endofunctor $\lift F$ on $\gcat{\R}{\cat{C}}$ lifting it, initial
$F$-algebras lift to initial $\lift F$-algebras.
\end{corollary}

\subsection{Final-coalgebra lifting theorem}

\begin{definition}
A $\lscat{C}$-indexed poset $\R$ is said to have 
\emph{existential quantification} whenever, for all $f: a\to b$ in $\lscat C$,
the monotone function $f\upstar:\R(b)\to\R(a)$ has a left adjoint, for which
we write $\lanexists f: \R(a)\to\R(b)$.
\end{definition}

\begin{lemma} \label{CoAlgebraKeyLemma}
For a $\lscat C$-indexed poset $\R$ with existential quantification, let $F$
be an endofunctor on $\lscat C$ and $\lift F$ be an endofunctor on
$\gcat{\R}{\cat{C}}$ lifting it. 

For a coalgebra $\delta: d\to Fd$, the greatest post-fixpoint
$\gpfp_\delta\in\R(d)$ of the monotone operator $\delta\upstar\comp\lift F_d$,
whenever it exists, provides a lifting of $\delta$ such that for all liftings
$R\in\R(c)$ of a coalgebra $\gamma: c\to Fc$, every homomorphism $c\to d$ from
$\gamma$ to $\delta$ lifts to a morphism 
$\Gobj c R \to \Gobj d {\gpfp_\delta}$.  
\end{lemma}
\begin{proof}
We have 
$\delta: \Gobj d {\gpfp_\delta} \to \Gobj {Fd\,}{\,\lift F_d(\gpfp_\delta)}$
because $\gpfp_\delta$ is a fixpoint of $\delta\upstar \comp \lift F_d$.

Let $\gamma: c\to Fc$ and $R\in\R(c)$ be such that 
$R \leq \gamma\upstar\big(\lift F_c(R)\big)$ and let $f: c \to d$ be an
homomorphism from $\gamma$ to $\delta$.

We prove $R \leq f\upstar(\gpfp_\delta)$ by equivalently showing 
$\lanexists f(R) \leq \gpfp_\delta$ establishing that $\lanexists f(R)$ is a
post-fixpoint of $\delta\upstar\comp\lift F_d$.  Indeed, 
\[\begin{array}{rcll}
R 
& \leq & \gamma\upstar\big(\lift F_c(R)\big) 
& \text{, by assumption}
\\[.5mm]
& \leq & \gamma\upstar(\lift F_c(f\upstar(\lanexists f(R))))
& \text{, as $\lanexists f\dashv f\upstar$}
\\[.5mm]
& \leq & \gamma\upstar((Ff)\upstar(\lift F_d(\lanexists f(R))))
& \text{, by Lemma~\ref{DirectConsequences}(\ref{LiftFOfCartesianLifting})}
\\[.5mm]
& = & f\upstar(\delta\upstar(\lift F_d(\lanexists f(R))))
& \text{, as $f:(c,\gamma)\to(d,\delta)$}
\end{array}\]
and therefore $\lanexists f(R) \leq \delta\upstar(\lift F_d(\lanexists f(R)))$.
\end{proof}

\begin{definition}
A \emph{co-ipo} is a poset whose opposite is an ipo.  A 
\emph{$\lscat C$-indexed co-ipo} is a \mbox{$\lscat C$-indexed} poset $\R$
such that $\R(c)$ is a co-ipo for all $c\in\lscat C$.
\end{definition}

By the dual of Pataria's theorem~\cite{Pataraia}, that monotone endofunctions
on co-ipos have greatest post-fixpoints, we obtain the following.

\begin{corollary}
For every $\lscat C$-indexed co-ipo with existential quantification $\R$,
every endofunctor $F$ on $\lscat C$, and every endofunctor $\lift F$ on
$\gcat{\R}{\lscat C}$ lifting it, final $F$-coalgebras lift to final
$\lift F$-coalgebras.
\end{corollary}

We note that the above may be also established under slightly weaker
hypothesis than existential quantification.

\begin{lemma} 
For a $\lscat C$-indexed poset $\R$, let $F$ be an endofunctor on $\lscat C$
and $\lift F$ be an endofunctor on $\gcat{\R}{\cat{C}}$ lifting it. 
For a coalgebra $\delta: d\to Fd$, such that $\R(d)$ is a co-ipo let
$\gpfp_\delta\in\R(d)$ be the greatest post-fixpoint of the monotone operator 
$\delta\upstar\comp\lift F_d$.

For a coalgebra $\gamma: \Gobj c R\to \Gobj{Fc}{\lift F_c(R)}$ and a
homomorphism $f: c\to d$ from $\gamma$ to $\delta$, if 
$\setof{\, S\in\R(d) \suchthat R\leq f\upstar(S) \,}$ is a sub co-ipo of
$\R(d)$ then $f$ lifts to a morphism $\Gobj c R \to \Gobj d {\gpfp_\delta}$.  
\end{lemma}
\begin{proof}
It suffices to show that $\setof{\, S\in\R(d) \suchthat R\leq f\upstar(S) \,}$
is invariant under $\delta\upstar\comp\lift F_d$. Indeed, if 
$R\leq f\upstar(S)$, then $\lift F_c(R)
  \leq \lift F_c\big(f\upstar(S)\big) 
  \leq (Ff)\upstar\big(\lift F_d(S)\big)$
and
$R \leq \gamma\upstar\big(\lift F_c(R)\big) 
   \leq \gamma\upstar\big((Ff)\upstar(\lift F_d(S))\big)
   = f\upstar\big(\delta\upstar\big(\lift F_d(S)\big)\big)$.
\end{proof}

\section{Focused orthogonality fibrationally}
\label{sec:focusedFib}
We study focused orthogonality categories representing them in terms of
Grothendieck categories of indexed complete lattices with existential
quantification.
This, together with the study of the previous section, provides an axiomatic theory of fixpoint constructions
in focused orthogonality models of linear logic.

\subsection{Focused orthogonality categories}\label{sec:focusedOrth}

We expand upon the exposition of focused orthogonality given in
Section~\ref{sec:prelim}.  A \emph{pole} in a category $\lscat C$ is a subset
$\Perp\ \subseteq \cat{C}(\source,\target)$ for a pair of distinguished
objects $\source$ and $\target$.  To obtain a model of intuitionistic linear
logic one takes $\source$ to be the monoidal unit $\One$, while in the
classical setting one further takes $\target$ to be its dual $\bot$.  The
\emph{focused orthogonality} induced by a pole $\Perp$ is the family of
relations 
$\setof{\, 
  \Perp_c \ \subseteq \cat{C}(\source,c) \times \cat{C}(c,\target) 
 \,}_{c\in\lscat C}$
given by 
\[ 
  x\Perp_c u \iff (u \comp x) \in \: \Perp
  \enspace.  
\]
The defining property of focused orthogonalities is being
\emph{reciprocal}~\cite{hamano2021double}; in the sense that, for all
morphisms $x : \source \rightarrow c$, $f : c \rightarrow d$, and 
$u: d \rightarrow \target$ in $\lscat C$, 
\begin{equation}\label{reciprocity}
  (f \circ x) \Perp_d u
  \iff 
	x \Perp_c (u \comp f) 
  \enspace.
\end{equation}
This plays a crucial role in the following section.

For a subset $X \subseteq \cat{C}(\source, c)$, its \emph{orthogonal}
$X^\orth \subseteq \cat{C}(c,\target)$ is given as below on the left
\[
  X^\orth 
  \eqdef 
  \setof{\, u : c \to \target \suchthat \forall\, x \in X.\, x \perp u \,}
  \enspace,
  \quad
  U^\orth 
  \eqdef
  \setof{\, x : \source \to c \suchthat \forall\, u \in U.\, x \perp u \,}
\]
while, dually, for a subset $U \subseteq \cat{C}(c,\target)$, its orthogonal
$U^\orth \subseteq \cat{C}(\source,c)$ is given as above on the right.
As it is standard, these definitions induce a Galois connection between $\C (c)\eqdef\big(\pow\big(\cat{C}(\source,c)\big),\subseteq\!\big)$
and $\Cd{c}\eqdef\big(\pow\big(\cat{C}(c,\target)\big),\supseteq\!\big)$.
The fixpoints of the associated closure operator on $\C (c)$, referred to as
\emph{double orthogonally closed} subsets, form complete lattices:
\begin{equation}\label{DoubleOrthogonallyClosed}
\D(c) 
\eqdef  
\setof{\, X \subseteq \cat{C}(\source, c) \suchthat X = X^\dorth \,}
\enspace.
\end{equation}

\begin{definition}
The \emph{focused orthogonality category} $\O_\Perp(\lscat C)$ of a category
$\lscat C$ with a pole ${\Perp\ \subseteq \cat{C}(\source,\target)}$ has objects
given by pairs $(c, X)$ with $c\in\lscat C$ and $X\in\D(c)$, and morphisms 
$f: (c, X)\to(d, Y)$ given by morphisms $f:c\to d$ in $\cat{C}$ such that
\begin{equation}\label{orthmorphism}
\forall\, (x:\source\to c) \in X.\, (f\comp x) \in Y 
\enspace.  
\end{equation}
\end{definition}

\subsection{Focused orthogonality categories are relational fibrations}

We need to introduce some notation.
\begin{enumerate}
\item
For a morphism $f : c\to d$ in a category $\cat{C}$, we respectively write
\[
\post{f} \eqdef \cat{C}(\source, f) : \cat{C}(\source, c) \to \cat{C}(\source,d)
\qquad \text{and} \qquad  	
\pre{f} \eqdef \cat{C}(f, \target) : \cat{C}(d, \target) \to \cat{C}(c,\target)  
\]
for the operations of post and pre composition with $f$. 

\item
For a function $h : A \to B$ between sets, we write 
$h\upstar : \pow(B) \to \pow(A)$ for the inverse image function and 
$\lanexists h : \pow(A) \to \pow(B)$ for its left adjoint.  In elementary
terms, 
\[
h\upstar(T) 
\eqdef 
\setof{\, {a\in A} \suchthat h(a)\in T \,} \quad \text{and} \quad \lanexists h (S) 
\eqdef 
\setof{\, b\in B \suchthat \exists\, a\in S.\, h(a)=b \,} \enspace .
\]
\end{enumerate}

For the rest of the section, let $f:c\to d$ be a morphism in a category
$\lscat C$ with a pole $\Perp\ \subseteq \cat{C}(\source,\target)$.  Since,
for $x:\source\to c$ in $\lscat C$ and $V\subseteq \lscat C(d,\target)$, 
\[
x\in(\post{f})\upstar(V^\orth)
\!\iff\! 
\forall\, v\in V.\, (f\comp x)\orth v
\quad\text{and}\quad
x\in\big(\lanexists{(\pre{\,f})}(V)\big)^\orth
\!\iff\! 
\forall\, v\in V.\, x\orth (v\comp f)
\]
we have 
\[
\big(\lanexists{(\pre{\,f})}(V)\big)^\orth = (\post{f})\upstar(V^\orth) 
\text{ for all } V\subseteq\lscat C(d,\target)
\]
and obtain the commutative diagram on the left below that recasts reciprocity
as a lifting property by duality:
\begin{equation} \label{AbstractReciprocityDiagram}
\begin{minipage}{.9\textwidth}\begin{center}
\begin{tikzpicture}[xscale=1.25,yscale=.66]
  \node (A) at (2,0) {$\C(c)$};
	\node (B) at (2,2) {$\Cd c$};
	\node (C) at (0,0) {$\C (d)$};
	\node (D) at (0,2) {$\Cd d$};
			
	\draw [->] (C) -- node [below] {$(\post{f})\upstar$} (A);
	\draw [->] (D) -- node [above] {$\lanexists{(\pre{\,f})}$} (B);
	\draw [->] (B) -- node [right] {$(-)^\orth$} (A);
	\draw [->] (D) -- node [left] {$(-)^\orth$} (C);
				
	\begin{scope}[xshift = 5cm]
	\node (A) at (2,0) {$\Cd c$};
	\node (B) at (2,2) {$\C (c)$};
	\node (C) at (0,0) {$\Cd d$};
	\node (D) at (0,2) {$\C (d)$};
						
	\node at (1,1) [rotate=-135] {$\Longrightarrow$};
	\draw [->] (C) -- node [below] {$\lanexists{(\pre{\,f})}$} (A);
	\draw [->] (D) -- node [above] {$(\post{f})\upstar$} (B);
  \draw [->] (B) -- node [right] {$(-)^\orth$} (A);
	\draw [->] (D) -- node [left] {$(-)^\orth$} (C);
	\end{scope}
\end{tikzpicture}
\end{center}\hfill\end{minipage}
\end{equation}

Then, as $(\post{f})\upstar:\C (d)\to\C (c)$ lifts along the right adjoints
$(-)^\orth$, it also lifts along the forgetful functors from their induced
categories of algebras; that is, in this case, it restricts to double
orthogonally closed subsets.
We spell out the details.

In~(\ref{AbstractReciprocityDiagram}), the diagram on the left has as mate the
diagram on the right; that is, 
\begin{center}
$\big((\post{f})\upstar(Y)\big)^\orth 
 \subseteq
 \lanexists{(\pre{\,f})}(Y^\orth)$
for all $Y\subseteq\lscat C(\source,d)$\enspace.
\end{center}
Pasting these two diagrams, we obtain a distributive law:
\begin{center}
  \begin{tikzpicture}[xscale=1.25,yscale=.66]
		
			\node (A) at (2,0) {$\Cd c$};
			\node (B) at (2,2) {$\C (c)$};
			\node (C) at (0,0) {$\Cd d$};
			\node (D) at (0,2) {$\C (d)$};
			
			\node at (1,1) [rotate=-135] {$\Longrightarrow$};
			\draw [->] (C) -- node [below] {$(\post{f})\upstar$} (A);
			\draw [->] (D) -- node [above] {$(\post{f})\upstar$} (B);
			\draw [->] (B) -- node [right] {$(-)^\dorth$} (A);
			\draw [->] (D) -- node [left] {$(-)^\dorth$} (C);
	\end{tikzpicture}
\end{center}

that is, $\big((\post{f})\upstar(Y)\big)^\dorth \subseteq (\post{f})\upstar(Y^\dorth)$
for all $Y\subseteq\lscat C(\source,c)$. Thus, $(\post{f})\upstar:\C (d)\to\C (c)$ restricts to a meet-preserving
monotone function $\D(d)\to\D(c)$ between complete lattices.

The above development results in a representation theorem for focused
orthogonality categories in terms of Grothendieck categories of indexed
complete lattices with existential quantification.

\begin{definition}
A \emph{$\lscat C$-indexed complete lattice} is a \mbox{$\lscat C$-indexed}
poset $\R$ such that $\R(c)$ is a complete lattice for all $c\in\lscat C$.
\end{definition}

\begin{theorem}
Every category $\lscat C$ with a pole $\Perp$ induces a $\lscat C$-indexed
complete lattice with existential quantification $\D$ whose Grothendieck
category $\gcat{\D}{\lscat C}$ is isomorphic to the focused orthogonality
category $\O_\Perp(\lscat C)$.
\end{theorem}
\begin{proof}
Consider the indexed family $\setof{\,\D(c)\,}_{c\in\lscat C}$ of double
orthogonally closed subsets~(\ref{DoubleOrthogonallyClosed}) with action
$\D(f)\eqdef(\post{f})\upstar:\D(d)\to\D(c)$ for all $f:c\to d$ in $\lscat C$.
Note that the condition $X\subseteq \D(f)(Y)$ is equivalent to the
statement~(\ref{orthmorphism}).
\end{proof}

\begin{corollary} \label{FocusedOrthogonalityLiftingCorollary}
Let $\lscat C$ be a category with a pole $\Perp$.  For every endofunctor $F$
on $\lscat C$ and every endofunctor $\lift F$ on $\O_\Perp(\lscat C)$ lifting
it, $F$ has an initial algebra (resp.~final coalgebra) if and only if so does
$\lift F$.
\end{corollary}

\section{Models of linear logic with fixpoints}
\label{Sec:Models}
\subsection{Categorical models}

We provide an alternative presentation of the notion of categorical model of
classical linear logic with fixpoints given by Ehrhard and
Jafarrahmani~\cite[Definition~7]{EJ2021}.  Our approach is adaptable to the
intuitionistic setting which we also include.  

We restrict attention to the specification of linear logic types; we omit the
specification of the logical system, the categorical models of which are well
known. The treatment of fixpoint operators requires the consideration of
types with variance in contexts of type variables with variance.  To this end,
we introduce judgements for types of the form $\Gamma \vdash \tau : v$ where
$v$ ranges over the set of sorts $\varsort \eqdef \setof{ + , - }$, $\Gamma$
ranges over $\varsort$-sorted contexts, and $\tau$ ranges over types.  The
sorts are used to indicate the intended variance, with $+$~(positive) standing
for covariance and $-$~(negative) standing for contravariance; as such the
dual sort $\dual v$ of a sort $v$ is given by $\dual+=-$ and $\dual-=+$.

The core judgement rules of the \emph{types of linear logic} 
are:\\[-3mm]
\[
\begin{array}{c}
\\ \hline
\Gamma\vdash x : v
\end{array}
\enspace(\text{$x:v$ in $\Gamma$})
\quad\qquad
\begin{array}{c}
  \\ \hline
  \Gamma\vdash\oper:+
\end{array}
\enspace\big(\text{$\oper$ in $\setof{\One,\Zero,\top}$}\big)
\]
\[
\begin{array}{c}
  \Gamma\vdash\tau_1:v
  \quad
  \Gamma\vdash\tau_2:v
  \\ \hline
  \Gamma\vdash\tau_1\oper\tau_2:v
\end{array}
\enspace\big(\text{$\oper$ in $\setof{\otimes,\oplus,\with}$}\big)
\quad\qquad
\begin{array}{c}
  \Gamma\vdash\tau:v
  \\ \hline
  \Gamma\vdash\oc\,\tau:v
\end{array}
\]
In classical linear logic, these are extended with:\\[-3mm]
\[
\begin{array}{c}
  \\ \hline
  \Gamma\vdash\bot:+
\end{array}
\enspace\qquad
\begin{array}{c}
  \Gamma\vdash\tau_1:v
  \quad
  \Gamma\vdash\tau_2:v
  \\ \hline
  \Gamma\vdash\tau_1\parr\tau_2:v
\end{array}
\enspace\qquad
\begin{array}{c}
  \Gamma\vdash\tau:v
  \\ \hline
  \Gamma\vdash\wn\,\tau:v
\end{array}
\enspace\qquad
\begin{array}{c}
  \Gamma\vdash\tau:v
  \\ \hline
  \Gamma\vdash\tau^\bot:\overline v
\end{array}
\]
while in intuitionistic linear logic they are extended 
with:\\[-3mm]
\[
\begin{array}{c}
  \Gamma\vdash\tau_1:\overline v
  \quad
  \Gamma\vdash\tau_2:v
  \\ \hline
  \Gamma\vdash\tau_1\multimap\tau_2:v
\end{array}
\]

A model of \emph{classical linear logic}
(resp.~\emph{intuitionistic linear logic}) is a $\star$-autonomous
(resp.\/ symmetric monoidal closed bicartesian) category with a linear
exponential
comonad~\cite{
seely1987linear,schalk2004categorical,mellies2009categorical}. 
For both classical and intuitionistic models $\lscat L$, every judgement
${x_1:v_1,\ldots,x_n:v_n}\vdash{\tau:v}$ has a standard \emph{interpretation}
functor $\lscat L^{v_1}\times\cdots\times\lscat L^{v_n}\to\lscat L^v$, where 
$\lscat L^+\eqdef\lscat L$ and $\lscat L^-\eqdef\lscat L^\op$.
The class of interpretation functors induced by judgements forms a
$\varsort$-sorted concrete clone $\mathcal L$ on $\lscat L$.

We recall the notion of \emph{parameterised fixpoint} (see, for instance,
Fiore~\cite[Chapter~6]{Fiore1996Axiomatic}).  A functor 
$F:\lscat D\times\lscat C\to\lscat C$ is said to have parameterised initial
algebras (resp.~final coalgebras) whenever, for all $d\in\lscat D$, the
endofunctor $F(d,-)$ on $\lscat C$ has an initial algebra (resp.~final
coalgebra), say $\mu F(d)$ \big(resp.~$\nu F(d)$\big), in which case the
structure canonically extends to a functor $\mu F:\lscat D\to\lscat C$
(resp.~$\nu F:\lscat D\to\lscat C$).

A $\varsort$-sorted concrete clone of functors 
$\mathcal F 
 = \setof{\,\mathcal F_{\sigma,v}\subseteq[\lscat C^\sigma,\lscat C^v]\,}
   _{\sigma\in\varsort^\star, v\in\varsort}$, 
where 
$\lscat C^{(s_1,\ldots,s_n)} 
 \eqdef \lscat C^{s_1}\times\cdots\times\lscat C^{s_n}$,
is said to have parameterised fixpoints whenever every 
$F\in\mathcal F_{(s_1,\ldots,s_n,v),v}$ has parameterised initial algebras and
final coalgebras, and their induced functors $\mu F$ and $\nu F$ are in 
$\mathcal F_{(s_1,\ldots,s_n),v}$.

A \emph{model of linear logic with fixpoints} is a model of linear logic
$\lscat L$ on which there is a \mbox{$\varsort$-sorted} concrete clone of
functors with parameterised fixpoints containing the $\varsort$-sorted clone 
$\mathcal L$ on $\lscat L$.
This notion of model is in line with the general notion of model for
second-order equational presentations~\cite{FioreHur} and allows for a
canonical interpretation of the extension of the linear logic typing
judgements with rules for least and greatest fixpoints as 
follows:\\[-2mm]
\[
\begin{array}{c}
  \Gamma,x : v \vdash \tau : v
  \\ \hline
  \Gamma \vdash \varphi\, x.\,\tau : v
\end{array}
\enspace(\text{$\varphi$ in $\setof{\mu,\nu}$})
\]

\subsection{Focused orthogonality models}

\begin{theorem}[Hyland and Schalk~\cite{GlueingHylandSchalk}]
\label{HSfocusedorthogonalitytheorem}
For a model of classical linear logic $\cat{L}$ with a distributive law 
$\cat{L}(\One,-) \to \cat{L}(\One,\oc -)$ and a pole 
$\Perp \: \subseteq \cat{L}(\One, \bot)$ between the monoidal units, 
the induced focused orthogonality category $\O_\Perp(\cat{L})$ is a model of
classical linear logic and the forgetful functor to $\cat L$ preserves the
structure strictly.
\end{theorem}
There is an analogous theorem for models of intuitionistic
linear logic for which the reader is referred to Hyland and
Schalk~\cite{GlueingHylandSchalk}.

The following result is a consequence of the theorem above and
Corollary~\ref{FocusedOrthogonalityLiftingCorollary}.
\begin{theorem}
\label{ModelLiftingTheorem}
Under the hypothesis of Theorem \ref{HSfocusedorthogonalitytheorem}, if
$\lscat L$ is a model of linear logic with fixpoints then so is
$\O_\Perp(\cat{L})$ and the forgetful functor to $\cat L$ preserves the
structure strictly. 
\end{theorem}

\subsection{Examples}\label{sec:examples}

A variety of models of linear logic are instances of focused orthogonality
constructions.  We examine examples to which Theorem~\ref{ModelLiftingTheorem}
applies and thereby yield models with fixpoints.  
As not all orthogonality models of linear logic are instances of focused
orthogonality constructions, we also discuss the cases of coherence and
finiteness spaces to which our results 
may be applied, even though these models do not arise from focused
orthogonalities in the relational model.

\begin{example} \big(Phase spaces~\cite{GirardPhase}\big) 
Consider a commutative monoid $(M, e, \cdot)$ and a subset 
$\Perp \: \subseteq M$.  For a subset $X \subseteq M$, its orthogonal is
defined as 
$X^\orth 
 \eqdef 
 \setof{\, y \in M \suchthat \forall x \in M, x \cdot y \in \Perp \,}$.
Subsets satisfying $X = X^\dorth$ are called \emph{facts} and provide a
complete provability semantics for linear logic.  
The commutative monoid $M$ can be considered as a compact closed category
$\cat{M}$ with a single object $\bullet$ (being both $\One$ and $\bot$).  The
pole $\Perp \: \subseteq \cat{M}(\bullet,\bullet)$ corresponds to a subset of
$M$ and one can reformulate the phase model within the setting of focused
orthogonality (except for the exponential structure which is defined
differently).  Therefore, one can interpret least and greatest fixpoints of
multiplicative and additive linear logic types in phase semantics to provide a
Tarskian sound model of $\MUMALL$~\cite{de2022phase}. 
\end{example}

The category $\Rel$ of sets and relations between them is one of the most
basic models of linear logic, with many other models arising as refinements of
it.  Being compact closed, it is a degenerate model.
The monoidal units are singletons and there are only two non-trivial focused orthogonalities on $\Rel$ given by the
poles $\{\,\varnothing\,\}$ and $\{\, \{ \id \} \, \} $.

\begin{example} 
The model of non-uniform totality spaces studied by Ehrhard and
Jafarrahmani~\cite{EJ2021} corresponds to the focused orthogonality category
$\Rel_{\Perp}$ induced by the pole $\Perp = \{\, \{ \id \} \, \}$.
Explicitly, for a set $A$ and a subset 
$X \subseteq \Rel(\One, A) \cong \pow(A)$,  one has:
\[
X^\orth 
	= \{\, y \in \pow(A) \mid \forall x \in X, y \circ x = \id \,\} 
	= \{\, y \in \pow(A) \mid \forall x \in X, x \cap y \neq \varnothing \,\}
 \enspace.
\]
The induced category is a model of $\MULL$ that provides a normalization
theorem for proofs.  
\end{example}

One can generalize the relational model by considering the category of
\emph{weighted relations} (or \emph{matrices}) $\Rel_{\Sring}$ over a continuous semiring $\Sring$
~\cite{Lamarche,laird2013weighted}.  
Standard examples are the Boolean semiring 
$(\{\true, \false\},  \vee, \wedge, \false, \true)$, the completed natural
numbers $\Ninfty= (\N \cup \{ \infty\},+, \cdot, 0,1)$, and the completed
non-negative reals $\Rinfty= (\Rp\cup \{ \infty\},+, \cdot, 0,1)$.  
Objects of $\Rel_{\Sring}$ are sets and a morphism from $A$ to $B$ is a
function $f : A \times B \to \Sring$ (also called an $\Sring$-matrix). The
composite of $f : A \times B \to \Sring$ and $g: B \times C \to \Sring$ is the
function $A \times C \to \Sring$ given by 
$(g\comp f)(a,c)\eqdef\sum_{b\in B} f(a,b) \cdot g(b,c)$. Since $\Rel_\Sring(\One,\bot) \cong \Sring$, a pole consists of a subset of
$\Sring$.

\begin{example}\label{PCohModelExample}
The model of probabilistic coherence spaces $\PCoh$ by Danos and
Ehrhard~\cite{danos2011probabilistic} can be obtained as a focused
orthogonality category on $\Rel_{\Rinfty}$ with pole 
$[0,1] \subseteq \Rinfty$.  The associated coKleisli category provides a fully
abstract model of probabilistic PCF~\cite{ehrhard2014probabilistic} with a
fixpoint operator for terms that may be obtained as an application of our
results in Section~\ref{sec:enrichedterms}~(see
Example~\ref{PCohFixpointExample}).
	
Hamano~\cite{hamano2021double} considered a continuous extension of $\PCoh$
based on the category of measurable spaces and s-finite transition kernels.
Even if not providing monoidal closed structure, his construction involves a
focused orthogonality.  Indeed, taking the same pole 
$\Perp\, := [0,1] \subseteq \Rinfty$, for a measurable space $(A, \Sigma)$, a
measure $\mu$ viewed as a morphism $(\One, \pow \One) \to (A, \Sigma)$, and a
measurable function $f$ viewed as a morphism 
$(A, \Sigma) \to (\One, \pow \One)$, one has $\mu \Perp_{(A,\Sigma)} f$ iff
$\int_A f \, d\mu \leq 1$.
\end{example}

\begin{example}
Orthogonality can be also used to relate models.  For instance, by considering
the qualitative linear logic model $\ScottL$, whose objects are preorders and
morphisms are ideal binary relations, the category of preorders and
projections introduced by Ehrhard~\cite{EhrhardScottRelCollapse} can be
obtained as a subcategory of the focused orthogonality model 
$(\ScottL \times \Rel)_{\Perp}$ with pole $\Perp := \{\, (\id, \id) \,\}$.  A
reflexive object in this setting, obtained by lifting reflexive objects from
$\ScottL$ and $\Rel$, allows the translation of normalization theorems between
idempotent and non-idempotent typing systems~\cite{ehrhard2012collapsing}.
\end{example}

\begin{example}
Coherence spaces, first introduced by Girard~\cite{GirardCoherence} to give a
denotational semantics for System~$\SF$, led to the discovery of linear logic
through the linear decomposition of stable functions.  The category of
coherence spaces $\Coh$ can be obtained as an orthogonality 
construction 
on $\Rel$: two subsets of a set are orthogonal whenever their intersection has
cardinality at most one. 
	
The orthogonality for coherence spaces in $\Rel$ is not focused; however, in
$\Coh$ one can consider the refinement of \emph{coherence spaces with
totality} that corresponds to the focused orthogonality category
$\Coh_{\Perp}$ with pole 
$\Perp\, := \{\, \{ \id \} \,\} \subseteq \Coh(\One,\bot)$.  
\end{example}

\begin{example} 
The model of (differential) linear logic of finiteness spaces by
Ehrhard~\cite{ehrhardFiniteness} is based on an orthogonality in $\Rel$ that
captures finite computations: two subsets of a set are orthogonal whenever
their intersection is finite.
This is however not focused and thus one cannot directly apply the results
developed in the paper.
Tasson and Vaux~\cite{tasson2018transport} studied conditions for lifting
endofunctors on $\Rel$ to $\Fin$ and showed how to calculate least fixpoints
for a subclass of linear logic formulas.  It remains to be seen whether
finiteness spaces provide a model for $\MULL$.
Instead, one can extend the model by considering a weighted version of
finiteness spaces over $\Rel_{\Ninfty}$ with pole 
$\Perp\, := \N \subseteq \Ninfty$, obtaining a focused model of linear logic
with fixpoints.
\end{example}

\section{Domain-theoretic models}
\label{sec:enrichedterms}
After Freyd~\cite{freyd1991algebraically}, the category-theoretic solution of
recursive type equations, where one is interested in fixpoints of recursive
types with mixed variance, is based on the notion of algebraic compactness
asserting the coincidence of initial algebras and final coalgebras.  
In domain-theoretic models, this in turn arises from the limit/colimit
coincidence in the order-enriched
setting~\cite{ScottContLat,Wand,SmythPlotkin1982}.

We write $\Cpo$ for the category of cpos ($\omega$-chain complete partial
orders) and continuous functions between them.

\begin{theorem}[{Fiore~\cite[Chapter~7]{Fiore1996Axiomatic}}]
Let a \emph{kind} be a $\Cpo$-category with ep-zero (viz.~a zero object such
that every morphism with it as source is an embedding) and colimits of
\mbox{$\omega$-chains} of embeddings.

Every kind is $\Cpo$-algebraically compact; that is, every $\Cpo$-endofunctor
on it has a bifree algebra (viz.~an initial algebra whose inverse is a final
coalgebra).  
\end{theorem}

In domain-theoretic models of linear logic, fixpoint operators arise
naturally and are typically charaterized by the axiom of uniformity.

\begin{definition}\label{def:uniffix}
\begin{enumerate}
\item
A \emph{fixpoint operator} on a category $\cat{D}$ with a terminal object
$\top$ is a family of functions 
$(-)^\dagger: \cat{D}(d,d) \to \cat{D}(\top, d)$ indexed by the objects $d$ of
$\cat{D}$ such that $f^\dagger = f \circ f^\dagger$ for all endomorphisms $f$
on $d$.  

\item
Let $\cat{C}$ and $\cat{D}$ be categories with terminal object, and let 
$J : \cat{C} \to \cat{D}$ be a bijective-on-objects functor preserving
terminal objects.  A fixpoint operator $(-)^\dagger$ on $\cat{D}$ is said to
be \emph{$J$-uniform} if for every $h: c \to d$ in $\cat{C}$ and 
$f :c \to c, g : d\to d$ in $\cat{D}$,
\[
J(h) \circ f = g \circ J(h) \quad \text{implies} \quad J(h) \circ f^\dagger= g^\dagger
\enspace.
\]
\end{enumerate}
\end{definition}

The following is a consequence of the study of fixpoint operators by Simpson
and Plotkin~\cite{simpson2000complete}. 

\begin{corollary} \label{UniformFixpointOperatorCorollary}
For a kind $\lscat D$ equipped with a $\Cpo$-comonad $S$ on it, the coKleisli
category $\lscat D_S$ has a unique $J$-uniform fixpoint operator for 
$J:\lscat D\to\lscat D_S$ the cofree functor of the adjoint resolution of
$S$.  
\end{corollary}

The above results apply, for instance, to the relational model and the coherence
space model (for the quantitative and qualitative comonads) of linear logic.  We now proceed to show that further examples
arise from Grothendieck categories in general and from focused orthogonality
in particular.

The following definition and theorem are instances of Section~5 of Cattani,
Fiore, and Winskel~\cite{CFWrecdom} where the categorified scenario is
considered. 
We will write $\Cppo$ for the full subcategory of $\Cpo$ consisting of the
pointed cpos (that is, those with bottom element) and let $\Cppo_\bot$ be the
subcategory of $\Cppo$ consisting of the strict (that is, bottom-element
preserving) functions.

\begin{definition}
An \emph{admissible $\lscat D$-indexed poset} for a $\Cppo_\bot$-category
$\lscat D$ is a $\Poset$-functor $\R: \lscat D^\op\to\Poset$ such that the
poset $\R(d)^\op$ is a cppo for all $d\in\lscat D$ and the monotone
function 
$\lscat D(c,d)\to\Poset(\R(d)^\op,\R(c)^\op): f\mapsto (f\upstar)^\op$ is
strict continuous for all $c,d\in\lscat D$.  
\end{definition}

\begin{theorem}
For a kind $\lscat D$ and an admissible $\lscat D$-indexed poset $\R$, the
Grothendieck category $\gcat\R{\lscat D}$ is a kind and the forgetful functor
to $\lscat D$ preserves the structure strictly.
\end{theorem}
\begin{proof}\hspace*{-1.25mm}\textbf{(outline)}\,\ 
Since $(\bigvee_n\,f_n)\upstar = \bigwedge_n\,(f_n)\upstar: \R(d)\to\R(c)$
for every $\omega$-chain $f$ in $\lscat D(c,d)$, the Grothendieck category
$\gcat\R{\lscat D}$ $\Cpo$-enriches.
Since, the reindexing $(\bot_{c,d})\upstar:\R(d)\to\R(c)$ along the bottom
element $\bot_{c,d}\in\lscat D(c,d)$ is constantly the top element of $\R(c)$,
the ep-zero of $\gcat\R{\lscat D}$ consists of the ep-zero $\bot$ of 
$\lscat D$ paired with the top element of $\R(\bot)$.
The colimiting cone $\seq{\, e_n : (d_n,R_n)\to(d,R) \,}_n$ of an
\mbox{$\omega$-chain} of embeddings 
$\seq{\, (d_n,R_n)\to(d_{n+1},R_{n+1}) \,}_n$ in $\gcat\R{\lscat D}$ consists
of the colimiting cone of embeddings $\seq{\, e_n : d_n\to d \,}_n$ of the
$\omega$-chain of embeddings $\seq{\, d_n\to d_{n+1} \,}_n$ in $\lscat D$ with 
$R \eqdef \bigwedge_n\,(p_n)\upstar(R_n)\in\R(d)$ for $p_n$ the
projection of $e_n$.
\end{proof}

Finally, we investigate focused orthogonality in this domain-theoretic
setting.

\begin{definition}
An \emph{admissible pole} for a $\Cppo_\bot$-category $\lscat D$ is a pole
that is a sub-cppo of $\lscat D(\source,\target)$.
\end{definition}

\begin{lemma}
For a $\Cppo_\bot$-category with an admissible pole, the indexed poset of
double orthogonally closed sets is admissible.
\end{lemma}

\begin{corollary}\label{cor:fixtermsenriched}
For a kind $\lscat D$ with an admissible pole $\Perp$, the focused
orthogonality category $\O_\Perp(\cat{D})$ is a kind and the forgetful functor
to $\lscat D$ preserves the structure strictly.
\end{corollary}

\begin{example} \label{PCohFixpointExample}
The weighted relational model $\Rel_{\Rinfty}$ (see
Example~\ref{PCohModelExample}) is a kind and the pole 
$\Perp\ \eqdef [0,1] \subseteq \Rinfty$ is admissible.
Corollaries~\ref{cor:fixtermsenriched}
and~\ref{UniformFixpointOperatorCorollary} provide then a uniform fixpoint
operator on the coKleisli category of probabilistic coherence spaces which
allows us to recover the fixpoint operator for terms 
of~\cite{danos2011probabilistic}.
	
On the other hand, totality models are tools to provide a denotational account
of normalization and therefore do not have fixpoint operators.  In particular,
note that for the totality models presented in Section~\ref{sec:examples} the
underlying orthogonality construction is done with the singleton pole
$\big\{\, \{ \id \} \, \big\}$ which does not contain the empty relation and
is therefore not admissible.  
\end{example}

\section*{Conclusion}

Recasting focused orthogonality constructions within a relational fibration
framework, we have developed a categorical theory to construct new models of
linear logic with fixpoints by means of lifting initial algebras and final
coalgebras from the base model to the focused orthogonality one.  
Our method is widely applicable: it allows to re-explain the totality model of
$\MULL$ studied by Ehrhard and Jafarrahmani~\cite{EJ2021} and opens the way
for refining a variety of other models besides the relational one. 
In connection to this, Tsukada and Asada~\cite{DBLP:conf/lics/TsukadaA22}
provided a unified framework based on module theory to make the linear
algebraic aspect of models of linear logic explicit.  In particular, they
considered models of intuitionistic linear logic based on categories of
$R$-modules and linear maps for $R$ a $\Sigma$-semiring.  
It would be interesting to investigate fixpoint constructions in these models
and thereafter consider refinements of them using our theory for focused
orthogonalities.
The same applies to their discussion of models of classical linear logic.

Our lifting theorems further extend from relational fibrations to categorical
fibrations.  In future work, we aim to use these results to obtain a theory of
fixpoint constructions for general glueing and double-glueing models. Since
double glueing constructions have been extensively used to study full
completeness by refining models to contain only morphisms that are the
interpretation of proof terms~\cite{loader1994linear}, we aim to also use our
results to construct fully complete models of linear logic with fixpoints. 

While we have considered fixpoint operators in the induced cartesian closed
category of domain-theoretic models of linear logic, we also aim to explore
lifting theorems for \emph{traces}~\cite{joyal1996traced} in the linear base
model.
Many orthogonality and (double) glueing constructions are indeed done on a
compact closed category (which has a canonical trace) and the refinement
induced by the orthogonality usually eliminates this degeneracy.
Understanding whether one can lift this canonical trace to orthogonality or
double-glued categories would provide a new method for constructing traced
categories for intuitionistic models.

\section*{Acknowledgements}
We are grateful to Thomas Ehrhard and Lionel Vaux Auclair for valuable discussions about this work.

\bibliographystyle{./entics}
\bibliography{biblio}

\end{document}